\documentclass[10pt,conference]{IEEEtran}
\usepackage{amsmath,graphicx,amssymb,amsfonts,bm}
\usepackage{cite,comment}
\usepackage{algorithm}
\usepackage{algorithmic}
\usepackage{balance}
\usepackage{ntheorem}

\usepackage[usenames, dvipsnames]{color}
 \newcommand{\rev}[1]{#1}
 \usepackage{tikz}
\usetikzlibrary{calc}
\usepackage{xcolor}
\usepackage{tabularx}
\usepackage{colortbl}
\usepackage{pgfplots}
\pgfplotsset{compat=newest}
\usetikzlibrary{plotmarks}
\usetikzlibrary{arrows.meta}
\usepgfplotslibrary{patchplots}

\usepackage[nolist]{acronym}
\begin{acronym}[ACRONYM]
\acro{AI}{artificial intelligence}
\acro{AoA}{angle-of-arrival}
\acro{AoD}{angle-of-departure}
\acro{BS}{base station}
\acro{BP}{belief propagation}
\acro{CDF}{cumulative density function}
\acro{CFO}{carrier frequency offset}
\acro{CPP}{carrier phase positioning}
\acro{CRB}{Cram\'er-Rao bound}
\acro{DA}{data association}
\acro{D-MIMO}{distributed multiple-input multiple-output}
\acro{DL}{downlink}
\acro{EKF}{extended Kalman filter}
\acro{EM}{electromagnetic}
\acro{FIM}{Fisher information matrix}
\acro{FR}{frequency range}
\acro{GDOP}{geometric dilution of precision}
\acro{GNSS}{global navigation satellite system}
\acro{GPS}{global positioning system}
\acro{IP}{incidence point}
\acro{IQ}{in-phase and quadrature}
\acro{ISAC}{integrated sensing and communication}
\acro{ICI}{inter-carrier interference}
\acro{JCS}{Joint Communication and Sensing}
\acro{JRC}{joint radar and communication}
\acro{JRC2LS}{joint radar communication, computation, localization, and sensing}
\acro{IMU}{inertial measurement unit}
\acro{IOO}{indoor open office}
\acro{IoT}{Internet of Things}
\acro{IRN}{infrastructure reference node}
\acro{KPI}{key performance indicator}
\acro{LoS}{line-of-sight}
\acro{LS}{least-squares}
\acro{LLF}{log-likelihood function}
\acro{MCRB}{misspecified Cram\'er-Rao bound}
\acro{MICRB}{mixed-integer Cram\'er-Rao bound}
\acro{MIMO}{multiple-input multiple-output}
\acro{ML}{maximum likelihood}
\acro{mmWave}{millimeter-wave}
\acro{NLoS}{non-line-of-sight}
\acro{NR}{new radio}
\acro{OFDM}{orthogonal frequency-division multiplexing}
\acro{OTFS}{orthogonal time-frequency-space}
\acro{OEB}{orientation error bound}
\acro{PEB}{position error bound}
\acro{PPP}{precise point positioning}
\acro{VEB}{velocity error bound}
\acro{PRS}{positioning reference signal}
\acro{QoS}{Quality of Service}
\acro{RAN}{radio access network}
\acro{RAT}{radio access technology}
\acro{RCS}{radar cross section}
\acro{RedCap}{reduced capacity}
\acro{RF}{radio frequency}
\acro{RIS}{reconfigurable intelligent surface}
\acro{RFS}{random finite set}
\acro{RMSE}{root mean squared error}
\acro{RTK}{real-time kinematic}
\acro{RTT}{round-trip-time}
\acro{SLAM}{simultaneous localization and mapping}
\acro{SLAT}{simultaneous localization and tracking}
\acro{SNR}{signal-to-noise ratio}
\acro{ToA}{time-of-arrival}
\acro{TDoA}{time-difference-of-arrival}
\acro{TR}{time-reversal}
\acro{TXRX}[TX/RX]{transmitter/receiver}
\acro{Tx}{transmitter}
\acro{Rx}{receiver}
\acro{UAV}{unmanned aerial vehicle}
\acro{UE}{user equipment}
\acro{UL}{uplink}
\acro{UWB}{ultra wideband}
\acro{XL-MIMO}{extra-large MIMO}

\acro{PL}{protection level}
\acro{TIR}{target integrity risk}
\acro{IR}{integrity risk}
\acro{RAIM}{receiver autonomous integrity monitoring}
\acro{1D}{one-dimensional}
\acro{3D}{three-dimensional}
\acro{SS}{solution separation}
\acro{PMF}{probability mass function}
\acro{PDF}{probability density function}
\acro{GM}{Gaussian mixture}
\acro{CCDF}{complementary cumulative distribution function}
\acro{RS}{reference station}

\end{acronym}
\newcommand{\Ptx}{P_{\text{tx}}}
\begin{document}
\bstctlcite{IEEEexample:BSTcontrol}
\title{Fundamental Performance Bounds for Carrier Phase Positioning in  Cellular Networks}
\author{ Henk~Wymeersch\IEEEauthorrefmark{1}, Rouhollah Amiri\IEEEauthorrefmark{2}, Gonzalo~Seco-Granados\IEEEauthorrefmark{3}, \\
 \IEEEauthorrefmark{1}Chalmers University of Technology, Gothenburg, Sweden,
 \IEEEauthorrefmark{2}Sharif University, Tehran, Iran,\\
 \IEEEauthorrefmark{3}Universitat Autonoma de Barcelona, Barcelona, Spain\\
 email: \tt{henkw@chalmers.se}, \tt{amiri\_rouhollah@ee.sharif.edu}, \tt{Gonzalo.Seco@uab.cat}
}
    
\maketitle
\begin{abstract}
    The carrier phase of cellular signals can be utilized for highly accurate positioning, with the potential for orders-of-magnitude performance improvements compared to standard time-difference-of-arrival positioning. Due to the integer ambiguities, standard performance evaluation tools such as the Cram\'er-Rao bound (CRB) are overly optimistic. In this paper, a new performance bound, called the mixed-integer CRB (MICRB) is introduced that explicitly accounts for this integer ambiguity. While computationally more complex than the standard CRB, the MICRB can accurately predict positioning performance, as verified by numerical simulations, {and hence it serves as a useful guide to choose the system parameters that facilitate carrier phase positioning.}
\end{abstract}
\section{Introduction}

In the evolution from 5G to 5G advanced and ultimately 6G, positioning has come more and more into focus \cite{nikonowicz_indoor_2022}. This is due to two compounding effects: a technology push and a requirements pull. The push is driven by the utilization of higher frequency bands with more available spectrum, the need for larger arrays at both the user and infrastructure side to overcome path loss, and the introduction of novel hardware (e.g., \ac{RIS}), novel deployments (e.g., cell-free MIMO) and novel methodologies (e.g., \ac{AI}). Combined, they will provide orders-of-magnitude positioning improvements compared to previous generations and enable new functionalities, such as radar-like sensing \cite{wild2021joint}. Complementary to this, the pull from the requirements leads to more strict demands regarding the relevant \acp{KPI}, including accuracy, latency, availability, and integrity, in support of use cases such as extended reality in autonomous robotics \cite{behravan2022positioning}.

Positioning quality is fundamentally tied to the quality of the underlying measurements, which typically include time and angle-based measurements, such as \ac{ToA}, \ac{AoA}, and \ac{AoD} \cite{wymeersch2022radio}. These measurements, in turn, involve the estimation of phase differences across subcarriers or antennas, with respect to arbitrary absolute phase references \cite{del2017survey}. Since the absolute phase of the signal is related to the propagation distance between the transmitter and the receiver, it can also be utilized in positioning, a process referred to as \ac{CPP} \cite{fouda_toward_2022}. 
\begin{figure}
    \centering
    \includegraphics[width=0.75\linewidth]{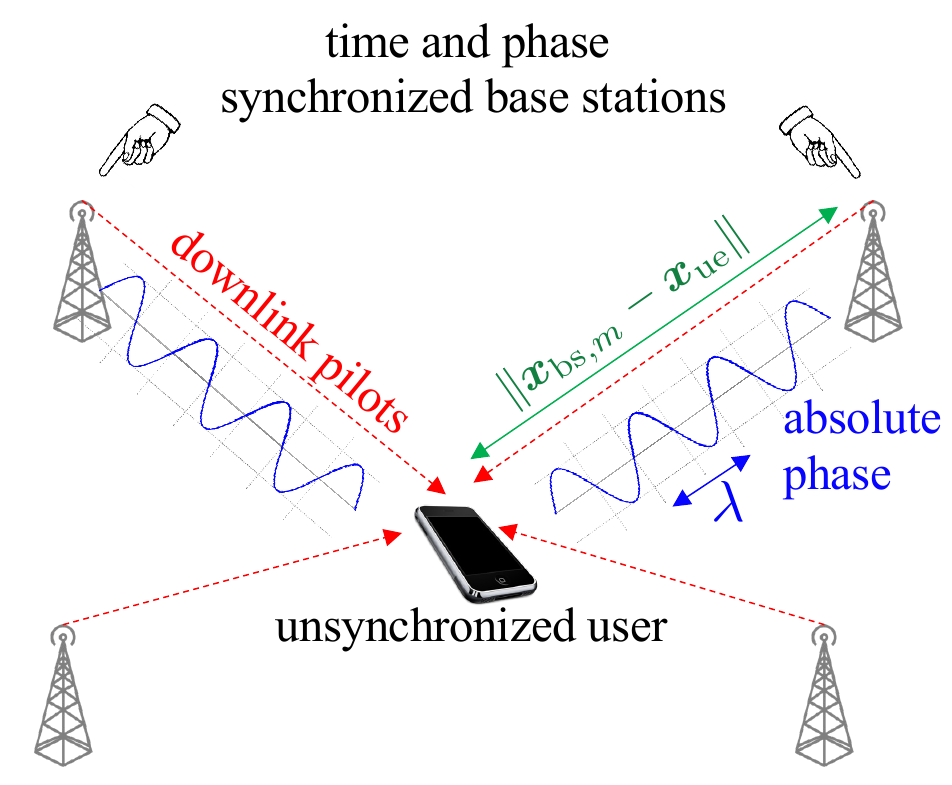}
    \caption{The received signal from each \ac{BS} information contains information about the distance in both the \ac{ToA} and the carrier phase. The \acp{BS} are time and phase synchronized, while the user is not.}
    \label{fig:CPPmodel}
\end{figure}
\Ac{CPP} has been studied extensively in the context of \ac{GNSS} positioning, via either \ac{PPP} or \ac{RTK}, relying on observations over time from several satellites \cite{teunissen2015review}. Inherent to all \ac{CPP} approaches is the so-called integer ambiguity, which refers to the fact that an observed phase is proportional to  distance modulo the signal wavelength. The integer ambiguity renders both the positioning problem, as well as its analysis, more challenging.
In the context of 5G, studies have considered both the integration with \ac{GNSS} \ac{CPP} \cite{zheng20235g} and 5G stand-alone \ac{CPP} \cite{fouda_toward_2022}. Focusing on the stand-alone operation, several studies have been conducted in recent years 
\cite{khalife2019opportunistic,abdallah2021uav,dun2020positioning,chen2021carrier,fascista2023uplink,fan2021carrier,fouda_toward_2022,li_carrier_2022,li_massive_2019}, considering both \ac{FR} 1 (below 6 GHz) and to a lesser extent \ac{FR}2 (mmWave, above 24 GHz). In \cite{nikonowicz_indoor_2022}, several practical challenges for \ac{CPP} are identified, including the integer ambiguities and multipath, and new opportunities are highlighted, including novel pilots optimized for \ac{CPP}. In FR1, \cite{khalife2019opportunistic,abdallah2021uav} demonstrated opportunistic 5G \ac{CPP} for \ac{UAV} tracking, while the other studies focused on simulations. In particular, \cite{dun2020positioning} shows how to perform carrier phase tracking for positioning with \ac{OFDM} signals from time and frequency synchronized \acp{BS}, considering aspects of multipath and phase wrapping, while \cite{chen2021carrier} shows that \ac{CPP} in FR1 provides sub-meter accuracies for both static and dynamic users. The use of reference stations, common in \ac{GNSS} \ac{RTK}, was also explored for 5G positioning \cite{fouda_toward_2022,li_carrier_2022}. 
\Ac{CPP} has also been combined with various MIMO configurations, such as massive MIMO 
\cite{li_massive_2019} for tracking phase information from multipath components,   cell-free MIMO \cite{fascista2023uplink} for evaluating the impact of dense multipath, and near-field tracking \cite{guerra2021near}. In FR2, \cite{fan2021carrier} uses \ac{CPP} for  inter-\ac{BS} synchronization and  \ac{UE} tracking. 
An in-depth performance evaluation of \ac{CPP} can be found in 3GPP report 
\cite{TR-38.859}. 
Surprisingly, despite this variety of algorithmic work, few papers have addressed the problem of fundamental performance bounds, which are useful for gaining deeper insights without the need for time-consuming simulations. While
\cite{vukmirovic2021performance,fascista2023uplink,guerra2021near} conducted \ac{CRB} analyses, neither considered the ambiguity. In \cite{medina2021cramer}, the  ambiguity was considered, but only in a small range. In \cite{guerra2021near}, the absolute phase was not utilized for positioning, so there was no ambiguity. 

{In this paper, we consider the \ac{CPP} problem in its simplest form (i.e., snapshot positioning of an unsynchronized \ac{UE} from \ac{ToA} and carrier phase measurements from  respect to several \acp{BS}, as shown in Fig.~\ref{fig:CPPmodel}) in order to derive a novel fundamental performance bound.} Our  contributions are as follows: (i) We develop the signal model, revealing when the integer ambiguity appears in going from the waveform observation to the delay and phase observation; (ii) We derive a novel bound, the \ac{MICRB}, which accounts for the integer ambiguity; (iii) Via simulations, we demonstrate the tightness and usefulness of the bound, and show under which conditions \ac{CPP} is most promising, including for cell-free MIMO systems.

\section{System Model}

We consider a scenario with 1 \ac{UE}, $M\ge 4$ \acp{BS}. The \ac{UE} has an unknown position $\bm{x}_{\text{ue}}\in \mathbb{R}^3$ and is not synchronized to the \acp{BS}. We further assume that the \acp{BS} are all mutually phase synchronized and have known locations $\bm{x}_{\text{bs},m}\in \mathbb{R}^3$. In case the time or phase synchronization does not hold, a \ac{RS} can be used for computing differential measurements in which the per-\ac{BS} time and phase biases are removed. 
 For simplicity, we assume that any frequency offsets among all entities have been corrected.

\Ac{OFDM} pilot signals with unit-modulus pilots are transmitted by the \acp{BS}, comprising $N$ subcarriers with subcarrier spacing $\Delta_f$. The $M$ \acp{BS} transmit orthogonal \ac{OFDM} waveforms (i.e., each \ac{BS} uses $N/M$ different subcarriers per OFDM symbol and $M$ OFDM symbols) with a transmit power $\Ptx$. After \rev{filtering, sampling, and cyclic prefix removal and combining over OFDM symbols,} the frequency-domain signal {(i.e., across the subcarriers)}  in terms of \ac{IQ} samples at  the \ac{UE} from \ac{BS} $m$ under ideal \ac{LoS} is given by~\cite[eq.~(1)--(2)]{wymeersch2022radio}
\begin{align}
    \bm{y}_{\text{iq},m}=\sqrt{E}_s\alpha_m \bm{d}(\tau_m) + \bm{w}_m,
\end{align}
where $E_s$ is the energy per subcarrier (with $E_s=\Ptx/(N \Delta_f)$), $\alpha_m =\rho_m e^{\jmath \vartheta_m}$ is the complex channel gain (in which $\rho_m \in \mathbb{R}$ captures the effect of path loss and transmitter and receiver antenna gains), $\tau_m$ is the \ac{ToA}, $[\bm{d}(\tau)]_n= e^{-{\jmath 2 \pi n \Delta_f \tau}}$ for $n\in \{0,\ldots,N-1\}-(N-1)/2$ is the centered delay steering vector, and $\bm{w}_m \sim \mathcal{CN}(\bm{0}_{N\times 1},N_0 \bm{I}_{N\times N})$ is the noise. We also introduce $W=N \Delta_f$ as the system bandwidth.
Under the stated assumptions, the \ac{ToA}  and carrier phase are related to the geometry by  
\begin{align}
    \tau_m & = \frac{1}{c} \Vert \bm{x}_{\text{bs},m}-\bm{x}_{\text{ue}} \Vert +B_{\text{ue}}\\
     \vartheta_m & =   \frac{2 \pi}{\lambda}  \Vert \bm{x}_{\text{bs},m}-\bm{x}_{\text{ue}} \Vert + \phi_{\text{ue}},
\end{align}
where  $B_{\text{ue}}$ is the clock bias of the  \ac{UE}, $\phi_{\text{ue}}$ is the phase bias of the \ac{UE}. 
The objective of the \ac{UE} is to estimate its position $\bm{x}_{\text{ue}}$ from $    \bm{y}_{\text{iq},m}$, for $m=1,\ldots, M$. 

\section{Observation Model and Performance Bounds}
In this section, we first present an observation model of the \ac{ToA} and carrier phase, from which we will derive the different bounds. 
\subsection{Intermediate Observation Model}
In order to derive all performance bounds, we will adopt a two-stage estimation process, where first the \ac{ToA} and carrier phase are estimated from $    \bm{y}_{\text{iq},m}$, as say $\hat{\tau}_{\text{ue},m}$ and $\hat{\vartheta}_{\text{ue},m}$, and then the UE position is determined. It is important to note that $\hat{\vartheta}_{\text{ue},m}$ is an estimate of $\phi_{\text{ue}} + {2 \pi}  \Vert \bm{x}_{\text{bs},m}-\bm{x} \Vert/\lambda$ modulo $2\pi$. Hence, we can express the \ac{ToA} and phase observations directly as distances $y_{\tau,m}=\hat{\tau}_{\text{ue},m}\times c$ and $y_{\vartheta,m}=\hat{\vartheta}_{\text{ue},m}\times \lambda/(2 \pi)$, given by 
\begin{align}
    y_{\tau,m} & =\Vert \bm{x}_{\text{bs},m}-\bm{x}_{\text{ue}} \Vert + B_{\text{ue}} c+ w_{\tau,m} \label{eq:y_tau}\\
    y_{\vartheta,m} &= \Vert \bm{x}_{\text{bs},m}-\bm{x}_{\text{ue}} \Vert + z_{m} \lambda + \phi_{\text{ue}}\frac{\lambda}{2 \pi}+ w_{\vartheta,m},\label{eq:y_theta}
\end{align}
where $z_{m} \in \mathbb{Z}$ is unknown.

Stacking observations from different \acp{BS}, we introduce $\bm{y}=[\bm{y}^\top_\tau \bm{y}^\top_\vartheta]^\top$, where $\bm{y}_\tau=[y_{\tau,1},\ldots,y_{\tau,M}]^\top \in \mathbb{R}^{M}$ and $\bm{y}_{\vartheta}=[y_{\vartheta,1},\ldots, y_{\vartheta,M}]^\top \in \mathbb{R}^M$. Stacking the unknowns, we introduce $\bm{\eta}=[\bm{s}^\top,\bm{z}^\top]^\top \in \mathbb{R}^{5}\times \mathbb{Z}^{M}$, where $\bm{s}=[\bm{x}^\top_{\text{ue}},B_{\text{ue}},\phi_{\text{ue}}]^\top$. As shown in the  Appendix, the lower bounds on the error covariance matrices of the noises $\bm{w}_\tau$ and $\bm{w}_\vartheta$ are diagonal matrices $\bm{\Sigma}_{\tau}$ and $\bm{\Sigma}_{\vartheta}$, where $[\bm{\Sigma}_{\tau}]^{-1}_m={2\,\text{SNR}_m  \pi^2 W^2}/{(3 c^2)}$ and $[\bm{\Sigma}_{\vartheta}]^{-1}_m={8\, \text{SNR}_m  \pi^2 }/{ \lambda^2}$, in which $\text{SNR}_m=N E_s\rho_m^2/N_0$. 

\subsection{Classical Performance Bounds} \label{sec:ClassicalBounds}
\rev{From the observation model \eqref{eq:y_tau}--\eqref{eq:y_theta}, classical performance bounds are derived, which ignore the integer ambiguities $\bm{z}$ \cite{van2004detection}, or relax the ambiguities to real numbers. 
}

\subsubsection{Known Integer Ambiguity}
When $\bm{z}$ is known, an optimistic bound is obtained. The \ac{FIM} of $\bm{s}=[\bm{x}^\top_{\text{ue}},B_{\text{ue}},\phi_{\text{ue}}]^\top$ is given by 
\begin{align} \label{eq:knownFIM}
  &   \bm{J}_{\text{known}}(\bm{s}) \\
 & =  \left[\begin{array}{ccc}
 \bm{U} \bm{J} \bm{U}^\top & c\bm{U}\text{diag}(\bm{J}_{\tau} )&  \frac{\lambda}{2\pi} 
\bm{U}\text{diag}(\bm{J}_{\vartheta} )  \\
c(\bm{U}\text{diag}(\bm{J}_{\tau} ))^\top & \text{tr}(\bm{J}_{\tau}) c^2 & 0\\
\frac{\lambda}{2\pi} 
(\bm{U}\text{diag}(\bm{J}_{\vartheta} ))^\top  & 0 & 
\text{tr}(\bm{J}_{\vartheta}) \frac{\lambda^2}{(2\pi)^2}
\end{array}\right],\notag 
\end{align}
where $\bm{J}_{\vartheta}=\bm{\Sigma}^{-1}_{\vartheta}$, $\bm{J}_{\tau}=\bm{\Sigma}^{-1}_{\tau}$, $\bm{J}=\bm{J}_{\vartheta}+\bm{J}_{\tau}$, and $ \bm{U}=[\bm{u}_1,\ldots ,\bm{u}_M]$, in which $\bm{u}_m=(\bm{x}_{\text{ue}}-\bm{x}_{\text{bs},m})/\Vert\bm{x}_{\text{ue}}-\bm{x}_{\text{bs},m} \Vert $. By inverting $\bm{J}_{\text{known}}(\bm{s})$, the error covariance bound on the position is readily obtained, i.e., $\bm{\Sigma}_{\text{known}}(\bm{x}_{\text{ue}})=[\bm{J}^{-1}_{\text{known}}({\bm{s}})]_{1:3,1:3}$.

\subsubsection{Delay-only Bound}
When the carrier phase observation $\bm{y}_{\vartheta}$ is not considered, the \ac{FIM} of $\tilde{\bm{s}}=[\bm{x}^\top_{\text{ue}},B_{\text{ue}}]^\top$ is given by 
\begin{align}
    \bm{J}_{\text{delay}}(\tilde{\bm{s}})= &  \left[\begin{array}{cc}
 \bm{U} \bm{J}_\tau \bm{U}^\top & c\bm{U}\text{diag}(\bm{J}_{\tau} )  \\
c(\bm{U}\text{diag}(\bm{J}_{\tau} ))^\top & \text{tr}(\bm{J}_{\tau}) c^2
\end{array}\right]. \label{eq:delay-onlyFIM}
\end{align}
After inversion and extraction of the first $3 \times 3 $ block, we obtain  $\bm{\Sigma}_{\text{delay}}(\bm{x}_{\text{ue}})=[\bm{J}^{-1}_{\text{delay}}(\tilde{\bm{s}})]_{1:3,1:3}$.

\rev{\subsubsection{Floating Integer Ambiguity Bound}
We consider $\bm{z}$ is an unconstrained real vector, so that in \eqref{eq:y_theta}, $\phi_{\text{ue}}/{(2 \pi)}$ is absorbed in each $z_m$. The estimate of  $\bm{z}\in \mathbb{R}^M$ is known as the float solution.  Hence, $\bm{\eta}$ becomes
$\bm{\eta}=[\tilde{\bm{s}}^\top,\bm{z}^\top]^\top \in \mathbb{R}^{4+M}$. After some manipulations of the \ac{FIM} of $\bm{\eta}$, it can be shown that 
the bound on the error covariance of $\bm{z}\in \mathbb{R}^M$ is
\begin{align}     \bm{\Sigma}_{\text{unc}}=\frac{1}{\lambda^2}\bm{\Sigma}_{\vartheta} + \frac{1}{\lambda^2}\bm{U}^\top \bm{\Sigma}_{\text{delay}}(\bm{x}_{\text{ue}}) \bm{U},
\end{align}
while the bound on the error covariance on $\tilde{\bm{s}}$ is the same as in the delay-only case.}

\subsection{Proposed Mixed-Integer Bound}
 We first express the stacked observation as 
\begin{align}
    \bm{y}& =\tilde{\bm{f}}(\tilde{\bm{s}}) + \bm{B}\bm{z}+ \kappa \bm{B}\bm{1}_{M\times 1}+ \bm{w}, \label{eq:completeObservation1}
\end{align}
where $\bm{z} \in \mathbb{Z}^M$,  $\bm{w}\sim \mathcal{N}(\mathbf{0}_{2M},\bm{\Sigma}_{\text{ch}})$, $\bm{\Sigma}_{\text{ch}}=\text{blkdiag}(\bm{\Sigma}_{\tau},\bm{\Sigma}_{\vartheta})$, 
$\bm{f}(\cdot)$ is a nonlinear function of $\tilde{\bm{s}}$, 
defined as, for $m=1,\ldots,M$,  
\begin{align}
    [\tilde{\bm{f}}(\tilde{\bm{s}})]_m & = \Vert \bm{x}_{\text{bs},m}-\bm{x}_{\text{ue}} \Vert + B_{\text{ue}}c\\
    [\tilde{\bm{f}}(\tilde{\bm{s}})]_{M+m} & =\Vert \bm{x}_{\text{bs},m}-\bm{x}_{\text{ue}} \Vert.
\end{align}
In addition, 
$\bm{B}=[\bm{0}_{M\times M}; \lambda \bm{I}_{M\times M}]\in \mathbb{R}^{2M \times M}$, and $\kappa = \phi_{\text{ue}}/(2\pi)$. Note that $\kappa$ and $\bm{z}$ are not jointly identifiable. To address this, we will reduce the state dimension. In particular, we express
\begin{align}
    \bm{y}=\tilde{\bm{f}}(\tilde{\bm{s}}) + \bm{B} 
    \left[\begin{array}{c}
         0  \\
         \bm{D} \bm{z} 
    \end{array}\right] +\kappa_{\text{d}} \bm{B}\bm{1}_{M\times 1} +\bm{w}, \label{eq:completeObservation2}
\end{align}
where $\bm{D}=[-\bm{1}_{(M-1)\times 1}\, \bm{I}_{(M-1)\times (M-1)}]$ and $\kappa_{\text{d}}=\kappa +z_1$, in which $z_1$ is the first\footnote{The choice of $\bm{D}$ is not unique provided that $\bm{D}\in \mathbb{Z}^{(M-1)\times M}$ satisfies $\bm{D}\bm{1}_{M\times 1}=\bm{0}_{(M-1)\times 1}$. It also can operate on a permutation of $\bm{z}$, say $\bm{P}\bm{z}$, where $\bm{P}\in \{0,1\}^{M\times M}$ is a permutation matrix. } entry of $\bm{z}.$ From this, we introduce $\bm{z}_{\text{d}}\doteq \bm{D} \bm{z} \in \mathbb{Z}^{M-1}$. 
Then we can write 
 \begin{align}
      \bm{y}=\bm{f}({\bm{s}}) + \bm{B} \bm{E}\bm{z}_{\text{d}} +\bm{w},\label{eq:completeObservation}
 \end{align}
 where $\bm{f}({\bm{s}})=\tilde{\bm{f}}(\tilde{\bm{s}})+\kappa_{\text{d}} \bm{B}\bm{1}_{M\times 1}$ and $\bm{E}=[\bm{0}^\top_{(M-1)\times 1}; \bm{I}_{(M-1)\times (M-1)}]$.

From the identifiable formulation \eqref{eq:completeObservation}, 
we proceed with \rev{the following steps: (i) Considering $\bm{z}\in \mathbb{R}^M$, we obtain  the floating integer ambiguity bound $\bm{\Sigma}_{\text{unc}}$ on $\bm{z}$ and the bound $\bm{\Sigma}_{\text{delay}}(\bm{x}_{\text{ue}})$ on 
$\bm{x}_{\text{ue}}$. 
(ii) We determine the probability of making specific errors $\bm{\delta}\in \mathbb{Z}^{M-1}$ when determining the integer solution $\bm{z}_{\text{d}}$ from the float solution.} (iii) An integer ambiguity error in  $\bm{z}_{\text{d}}$ leads to a biased estimate of $\bm{s}$. This bias will be characterized. (iv) We put everything together to compute the final error covariance on $\bm{s}$, considering the integer ambiguity errors, their biases, and their probabilities. 



\subsubsection{Integer Ambiguity Errors}
\rev{We decompose the float solution $    \hat{\bm{z}}_{\text{unc}}\in \mathbb{R}^M$ into}
    \begin{align}
    \hat{\bm{z}}_{\text{unc}} = \bm{E} \bm{z}_{\text{d}} + \kappa_{\text{d}} \bm{1}_{M\times 1}+ \bm{u}, \bm{u}\sim \mathcal{N}(\bm{0},\bm{\Sigma}_{\text{unc}}). \label{eq:floatsolution}
\end{align}
 We next determine a differential observation 
\begin{align}
    \bm{D}\hat{\bm{z}}_{\text{unc}} &  = \bm{D}\bm{E}\bm{z}_{\text{d}}+ \kappa_{\text{d}}\bm{D}\bm{1}_{M\times 1} +\bm{D}\bm{u}\\ 
    & = \bm{z}_{\text{d}}+ \bm{D}\bm{u}
\end{align} 
since $\bm{D}\bm{1}_{M\times 1}=\bm{0}_{(M-1)\times 1}$. We whiten the noise, leading to an observation $\bm{r}=\bm{S}^{-1/2} \bm{z}_{\text{d}} + \mathbf{u}'$, where $\bm{r}=\bm{S}^{-1/2}\bm{D}\hat{\bm{z}}_{\text{unc}}$,  $\mathbf{u}' \sim \mathcal{N}(\bm{0}_{(M-1)\times 1},\bm{I}_{(M-1)\times (M-1)})$ and $\bm{S}=\bm{D}\bm{\Sigma}_{\text{unc}}\bm{D}^\top$. Finally, we 
 recover $\bm{z}_{\text{d}}$ by solving the following integer problem 
\begin{align}
    \hat{\bm{z}}_{\text{d}}=\arg \min_{\bm{z}_{\text{d}}\in \mathbb{Z}^{M-1}} \Vert \bm{r} -\bm{S}^{-1/2} \bm{z}_{\text{d}}\Vert, \label{eq:estimationILS}
\end{align}
which can be solved efficiently with standard toolboxes \cite{chang2011miles}, provided $M$ is not too large. 
From $\hat{\bm{z}}_{\text{d}}$, we introduce the integer error $\bm{\delta}= \hat{\bm{z}}_{\text{d}}-\bm{z}_{\text{d}}$.

\subsubsection{Bias in $\bm{s}$ due to  Integer Ambiguity Error}
To determine the bias in ${\bm{s}}$, we linearize \eqref{eq:completeObservation} around ${\bm{s}}$. With a slight abuse of notation, this leads to $\bm{y}=\bm{A} {\bm{s}}+ \bm{B}\bm{E}\bm{z}_{\text{d}}+\bm{w}$, after removing irrelevant terms, and in which $\bm{A}=\nabla_{{\bm{s}}} \bm{f}({\bm{s}})\in \mathbb{R}^{2M\times 5}$, computed at the true value of $\bm{s}$. After whitening the noise and computing the least-squares estimate of ${\bm{s}}$, the bias is immediately recovered as
\begin{align}
    \bm{b}({\bm{s}}|\bm{\delta})= (\bm{\Sigma}^{-1/2}_{\text{ch}}\bm{A})^\dagger \bm{\Sigma}^{-1/2}_{\text{ch}} \bm{B}\bm{E}\bm{\delta},
\end{align}
where $\bm{X}^\dagger$ is the Moore-Penrose inverse of the tall matrix $\bm{X}$, i.e., $\bm{X}^\dagger=(\bm{X}^\top \bm{X})^{-1}\bm{X}^\top$.

\subsubsection{Mixed Integer \ac{CRB}}
To now computed the \ac{MICRB}, we first recall that when the estimator of $\tilde{\bm{s}}$ is biased with bias $\bm{b}({\bm{s}}|\bm{\delta})$, the resulting error covariance is (including the bias) \cite{eldar2006uniformly}
\begin{align}
    \bm{\Sigma}({\bm{s}}|\bm{\delta})=\bm{b}({\bm{s}}|\bm{\delta}) (\bm{b}({\bm{s}}|\bm{\delta}))^\top + (\bm{I}+\bm{H}_{\bm{b}})\bm{\Sigma}_{\text{known}}({\bm{s}})(\bm{I}+\bm{H}_{\bm{b}})^\top
\end{align}
where\footnote{In case the bias is not very sensitive to $\bm{s}$, $\bm{H}_{\bm{b}}\approx \bm{0}$, so that the error covariance is approximated as 
    $\bm{\Sigma}({\bm{s}}|\bm{\delta})\approx \bm{b}({\bm{s}}|\bm{\delta}) (\bm{b}({\bm{s}}|\bm{\delta}))^\top + \bm{\Sigma}_{\text{known}}({\bm{s}}).$
} $\bm{H}_{\bm{b}}=\nabla_{{\bm{s}}}\bm{b}({\bm{s}}|\bm{\delta}) \in \mathbb{R}^{5\times 5}$.
Finally, the proposed bound on the error covariance is obtained by taking the expectation with respect to the integer ambiguity error:
\begin{align}\label{eq:proposedbound}
    \bm{\Sigma}_{\text{mi}}({\bm{s}}) & = \sum_{\bm{\delta}\in \mathbb{Z}^{M-1}} \text{Pr}(\bm{\delta})\bm{\Sigma}({\bm{s}}|\bm{\delta}).
\end{align}
\rev{The summation can be efficiently computed via Monte Carlo simulation. 
To do this,  we generate $N_s$ samples $\bm{r}^{(i)}=\bm{S}^{-1/2} \bm{z}_{\text{d}} + \bm{u}^{(i)}$, where ${\bm{u}}^{(i)}\sim \mathcal{N}(\bm{0}_{M-1},\bm{I}_{M-1})$ and determine the estimate $\hat{\bm{z}}_{\text{d}}^{(i)}$. 
From this, we obtain ${\bm{\delta}}^{(i)}=\hat{\bm{z}}_{\text{d}}^{(i)}-\bm{z}_{\text{d}}$. Finally, $   \bm{\Sigma}_{\text{mi}}({\bm{s}})\approx {1}/{N_s}\sum_{i=1}^{N_s} \bm{\Sigma}({\bm{s}}|\bm{\delta}^{(i)})$.}

\rev{
\subsubsection{Complexity Analysis}
Due to the involved sampling (with complexity scaling linearly in $N_s$) and the need to solve \eqref{eq:estimationILS} for each sample (with complexity scaling at least cubically in $M$
\cite{hassibi2005sphere}), the complexity of the proposed \ac{MICRB} is far higher than the conventional counterparts from Section \ref{sec:ClassicalBounds}. 
}
\section{Positioning Algorithms}
In this section, we describe two distinct approaches to solve the \ac{CPP} problem: the first one is based directly on \eqref{eq:completeObservation}, while the second one is based on directional statistics. {These algorithms are not the main contributions of this work and are presented to provide a baseline for the bounds.}

\subsection{Mixed Integer Approach}\label{sec:mixedInteger}
The \ac{ML} problem can be expressed as 
\begin{align}
    \hat{\bm{s}},\hat{\bm{z}}_{\text{d}}=\arg \min_{\bm{s}\in \mathbb{R}^5,\bm{z_\text{d}} \in \mathbb{Z}^{M-1}} \Vert \bar{\bm{y}}-\bm{\Sigma}^{-1/2}_{\text{ch}}\bm{f}(\bm{s}) - \bar{\mathbf{B}}\bm{z}_{\text{d}}\Vert^2. \label{eq:mixedInteger}
\end{align}
where $\bar{\bm{y}}=\bm{\Sigma}^{-1/2}_{\text{ch}}\bm{y}$ and $\bar{\mathbf{B}}=\bm{\Sigma}^{-1/2}_{\text{ch}}\bm{B}\bm{E}$. Solving this problem is challenging, due to the combination of the nonlinearity $\bm{f}(\cdot)$ and the integer variable $\bm{z}_{\text{d}}$. To deal with this, the standard solution involves several steps \cite{wang_new_2021}. First, we
 linearize \eqref{eq:completeObservation} around some value ${\bm{s}}_0$:
\begin{align}
    \bm{y}\approx \bm{f}({\bm{s}}_0)+\nabla \bm{f}({\bm{s}}_0) \delta {\bm{s}}+ \bm{B}\bm{E}\bm{z}_{\text{d}} + \bm{w}, \label{eq:linearization}
\end{align}
where we recall that ${\bm{s}}=[\bm{x}^\top_{\text{ue}},B_{\text{ue}}, \phi_{\text{ue}}]^\top$. 
To obtain ${\bm{s}}_0$, we solve 
\begin{align}
    & \min_{\bm{x}_{\text{ue}},B_{\text{ue}}}  \sum_{m=1}^M \frac{\left(y_{\tau,m}- \Vert \bm{x}_{\text{bs},m}-\bm{x}_{\text{ue}} \Vert - c B_{\text{ue}} \right)^2}{2 \sigma^2_{\tau,m}} \label{eq:delayOnlyEstimate}
\end{align}
using any conventional \ac{TDoA} method. This provides ${\bm{s}}_0$, in which we set  $\hat{\phi}_{\text{ue}}=0$ since it appears linearly. 
Then, after removing $\bm{f}({\bm{s}}_0)$, we are left with 
\begin{align}
    \bm{y}' = \nabla \bm{f}({\bm{s}}_0) \delta {\bm{s}}+ \bm{B}\bm{E}\bm{z}_{\text{d}} + \bm{w}. \label{eq:linearized}
\end{align} 
We can then find an unconstrained estimate of $\bm{z}_{\text{d}}$ and then solve \eqref{eq:estimationILS} to obtain the integer estimate of 
$\bm{z}_{\text{d}}$. 
Finally, we obtain a closed-form solution of $\delta \bm{s}$ from \eqref{eq:linearized}, 
 considering $\bm{z}_{\text{d}}$ to be known. 

\subsection{Directional Statistics Approach}\label{sec:directional}

\rev{As an alternative approach, we can avoid integer ambiguities by working with directional statistics, where we model the carrier phase measurements with a von Mises distribution \cite{cai2007statistical}. Then, the  negative \ac{LLF}  becomes}
\begin{align}
    & -\log p(\bm{y}|\bm{x}_{\text{ue}},B_{\text{ue}},\phi_{\text{ue}}) = \label{eq:NLL}\\
    & \sum_{m=1}^M \frac{\left(y_{\tau,m}- \Vert \bm{x}_{\text{bs},m}-\bm{x}_{\text{ue}} \Vert - c B_{\text{ue}} \right)^2}{2 \sigma^2_{\tau,m}} + \notag\\
    &  \sum_{m=1}^M \frac{\lambda^2}{(2 \pi)^2\sigma^2_{\vartheta,m}} {\cos \Big(\Big( \frac{2 \pi}{\lambda } y_{\vartheta,m} -  \frac{2 \pi}{\lambda } \Vert \bm{x}_{\text{bs},m}-\bm{x}_{\text{ue}} \Vert\Big)-\phi_{\text{ue}}\Big)}.\notag
\end{align}
To solve \eqref{eq:NLL}, we note that for a given $\bm{x}_{\text{ue}}$, we can easily determine $B_{\text{ue}}$ (in closed form) and $\phi_{\text{ue}}$ (by a 1D search). 
The method thus proceeds as follows. From the \ac{ToA} measurements, a coarse estimate of $\bm{x}_{\text{ue}}$ is obtained, as in \eqref{eq:delayOnlyEstimate}. Then, a fine grid around $\bm{x}_{\text{ue}}$ is defined with step size dependent on the bound $\bm{\Sigma}_{\text{known}}(\bm{x}_{\text{ue}})$ and a grid size dependent on the bound $\bm{\Sigma}_{\text{delay}}(\bm{x}_{\text{ue}})$. For each trial location, $B_{\text{ue}}$ and $\phi_{\text{ue}}$ are determined and the solution with minimal negative log-likelihood is selected.  Overall, this method requires a fine 4D search, which is computationally complex.

\section{Numerical Results}

\subsection{Scenario}

We consider a UE at a fixed location $\bm{x}_{\text{ue}}=[0;0;5]^\top$ and $M$ BSs located in 3D space, with $\bm{x}_{\text{bs},m}\sim \mathcal{N}(\bm{0},(0.1~\text{km})^2\bm{I}_{3\times 3})$. The default system parameters are as follows:
carrier frequency $f_c=28~\text{GHz}$,  subcarrier spacing $\Delta_f=20~\text{kHz}$, $N=300$ subcarriers, $-174~\text{dBm/Hz}$ noise power spectral density, a receiver noise figure of $13~\text{dB}$, transmission power of $0~\text{dBm}$, and $M=7$ BSs. 
We set $\rho_m = \lambda / (4 \pi \Vert \bm{x}_{\text{bs},m}-\bm{x}_{\text{ue}} \Vert)$, to model the dependence of the \rev{free space} path loss on the carrier frequency. 

\subsection{Bounds and Algorithms}

Performance of algorithms will be measured in terms of the \ac{RMSE}, while the bounds will be represented by the \ac{PEB}, defined as $\text{PEB}=\sqrt{\text{trace}(\bm{\Sigma}(\bm{x}_{\text{ue}}))}$. The RMSE and PEB are both expressed in meters.  We will study the following bounds: the bound with known integer ambiguity\footnote{In all cases, this bound turned out to coincide with the generalized \ac{CRB} from \cite{medina2021cramer}, so the latter bound is not included in the results.} based on \eqref{eq:knownFIM}, the delay-only bound from \eqref{eq:delay-onlyFIM}, and the proposed \ac{MICRB} \eqref{eq:proposedbound} computed with $N_s=1000$. 
We will present the following algorithms: the mixed integer approach from Section \ref{sec:mixedInteger},  the directional statistics approach from Section \ref{sec:directional}, and the delay-only estimator. 

\subsection{Results and Discussion}

We will evaluate the impact of the following parameters: the carrier frequency, the bandwidth, the transmission power, and the number of BSs. When varying one parameter, the other parameters are set to their default values. 

\subsubsection{Impact of Carrier Frequency}
\begin{figure}[t]
    \centering    
    \definecolor{mycolor1}{rgb}{0.49400,0.18400,0.55600}%
\definecolor{mycolor2}{rgb}{0.63500,0.07800,0.18400}%
\begin{tikzpicture}[scale=1\columnwidth/10cm,font=\footnotesize]
\begin{axis}[%
width=8cm,
height=4cm,
scale only axis,
xmode=log,
xmin=1,
xmax=100,
xminorticks=true,
xlabel style={font=\color{white!15!black}},
xlabel={carrier frequency [GHz]},
ylabel style={font=\color{white!15!black}},
ylabel={[m]},
ymode=log,
ymin=0.0001,
ymax=10000,
yminorticks=true,
 legend columns=3,
 legend pos=north west,
axis background/.style={fill=white},
legend style={legend cell align=left, align=left, draw=white!15!black}
]
\addplot [color=blue, line width=1.0pt, dotted]
  table[row sep=crcr]{%
1.58489319246111	0.0898507135125\\
1.8873918221351	0.106999956024265\\
2.24762646922016	0.127422366964307\\
2.67661684547528	0.151742675474601\\
3.18748592597229	0.180704848832701\\
3.79586138578132	0.215194850687307\\
4.52035365636024	0.256267743015605\\
5.38312522557077	0.30517996086133\\
6.41056859642023	0.363427747149791\\
7.63411364353909	0.432792923315133\\
9.09118906472884	0.515397395880332\\
10.8263673387405	0.613768066366485\\
12.8927282139681	0.730914130149625\\
15.3534824376823	0.870419128865832\\
18.2839053962858	1.03655057228195\\
21.7736398173647	1.23439048300562\\
25.9294379740467	1.46999085744598\\
30.8784272767174	1.75055879863335\\
36.7719991477594	2.08467698417998\\
43.7904401414373	2.48256621357852\\
52.1484469766079	2.95639806625768\\
62.1016941891561	3.52066723472135\\
73.9546553110858	4.19263492258018\\
88.0699168290408	4.99285687118629\\
104.879269839711	5.94581216740209\\
124.896918699982	7.08065206796666\\
148.735210729351	8.43209174727536\\
177.123368142049	10.0414722474687\\
210.929795225637	11.9580251162786\\
251.188643150958	14.2403784183735\\
};
\addlegendentry{$\text{PEB}_{\text{delay}}$}

\addplot [color=black, line width=1.0pt, dashed,mark=o,mark options={solid, black},mark repeat=4]
  table[row sep=crcr]{%
1.58489319246111	0.000327309055805551\\
1.8873918221351	0.000327309696154826\\
2.24762646922016	0.000327310147693504\\
2.67661684547528	0.000327310466092985\\
3.18748592597229	0.000327310690610018\\
3.79586138578132	0.000327310848926423\\
4.52035365636024	0.000327310960561922\\
5.38312522557077	0.000327311039280739\\
6.41056859642023	0.000327311094788625\\
7.63411364353909	0.000327311133929519\\
9.09118906472884	0.000327311161529369\\
10.8263673387405	0.000327311180991155\\
12.8927282139681	0.000327311194714457\\
15.3534824376823	0.000327311204391318\\
18.2839053962858	0.000327311211214869\\
21.7736398173647	0.000327311216026434\\
25.9294379740467	0.000327311219419264\\
30.8784272767174	0.000327311221811688\\
36.7719991477594	0.000327311223498683\\
43.7904401414373	0.000327311224688253\\
52.1484469766079	0.000327311225527067\\
62.1016941891561	0.000327311226118549\\
73.9546553110858	0.000327311226535627\\
88.0699168290408	0.000327311226829726\\
104.879269839711	0.000327311227037107\\
124.896918699982	0.000327311227183339\\
148.735210729351	0.000327311227286454\\
177.123368142049	0.000327311227359164\\
210.929795225637	0.000327311227410436\\
251.188643150958	0.000327311227446589\\
};
\addlegendentry{$\text{PEB}_{\text{known}}$}

\addplot [color=red, line width=1.0pt, dashed]
  table[row sep=crcr]{%
1.58489319246111	0.000327309055805547\\
1.8873918221351	0.000327309696154831\\
2.24762646922016	0.000327310147693504\\
2.67661684547528	0.00032731046609299\\
3.18748592597229	0.000327310690610013\\
3.79586138578132	0.000327310848926426\\
4.52035365636024	0.000327310960561917\\
5.38312522557077	0.000327311039280732\\
6.41056859642023	0.000327311094788634\\
7.63411364353909	0.000327311133929522\\
9.09118906472884	0.000327311161529366\\
10.8263673387405	0.00032731118099116\\
12.8927282139681	0.000327311194714451\\
15.3534824376823	0.000327311204391317\\
18.2839053962858	0.000327311211214873\\
21.7736398173647	0.0878521506638894\\
25.9294379740467	0.293232601131285\\
30.8784272767174	0.638370109522929\\
36.7719991477594	1.48206007009735\\
43.7904401414373	2.21907875330068\\
52.1484469766079	3.05921890210669\\
62.1016941891561	3.43206197886956\\
73.9546553110858	4.00000384999616\\
88.0699168290408	4.70072035237706\\
104.879269839711	5.42656972519865\\
124.896918699982	6.48863647671514\\
148.735210729351	7.58133871900283\\
177.123368142049	8.62278694281623\\
210.929795225637	10.5709178649824\\
251.188643150958	12.1016759861581\\
};
\addlegendentry{$\text{PEB}_{\text{mi}}$}

\addplot [color=blue, line width=1.0pt, mark=square, mark options={solid, blue}]
  table[row sep=crcr]{%
1.58489319246111	0.0899111558942891\\
2.78255940220712	0.150557698916214\\
4.88527357151939	0.25289569287468\\
8.57695898590894	0.523009120828749\\
15.0583635427984	0.832302258105876\\
26.437611857491	1.49112253792572\\
46.4158883361278	2.6074189090738\\
81.4912746902074	4.23488595690528\\
143.072298919376	7.73748708395285\\
251.188643150958	13.3315382176314\\
};
\addlegendentry{$\text{RMSE}_{\text{delay}}$}

\addplot [color=red, line width=1.0pt, mark=o, mark options={solid, red}]
  table[row sep=crcr]{%
1.58489319246111	0.000338323816294988\\
2.78255940220712	0.000343159293335098\\
4.88527357151939	0.000339496700992849\\
8.57695898590894	0.000334254671217417\\
15.0583635427984	0.000354581091840855\\
26.437611857491	4.41176793831667\\
46.4158883361278	7.57337815113661\\
81.4912746902074	13.2366445532713\\
143.072298919376	22.0734910808662\\
251.188643150958	42.327347348547\\
};
\addlegendentry{RMSE \eqref{eq:NLL}}

\addplot [color=black, line width=1.0pt]
  table[row sep=crcr]{%
1.58489319246111	0.000330964669364824\\
2.78255940220712	0.000344602136408618\\
4.88527357151939	0.000362581166078667\\
8.57695898590894	0.000589717606941722\\
15.0583635427984	1.15299562009147\\
26.437611857491	2.19476303017006\\
46.4158883361278	3.19393382135539\\
81.4912746902074	4.86050841366769\\
143.072298919376	7.74836208718489\\
251.188643150958	13.657807480192\\
};
\addlegendentry{RMSE \eqref{eq:mixedInteger}}

\end{axis}
\end{tikzpicture}%
   \vspace{-10mm}
    \caption{Impact of carrier frequency on PEB and positioning RMSE.}
    \label{fig:carrier}
\end{figure}
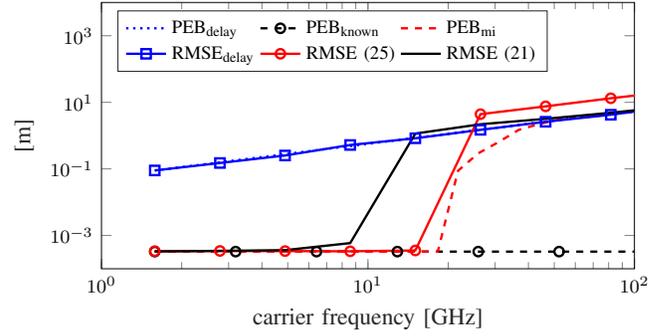

\begin{figure}[t]
    \centering
    \include{Figures/NLL-3GHzL}
    \vspace{-10mm}
    \caption{\Ac{LLF} at 3 GHz carrier for delay only and combined delay and carrier phase observations.}
    \label{fig:NLL3GHz}
\end{figure}

\begin{figure}[t]
    \centering
    \include{Figures/NLL-30GHzL}
    \vspace{-10mm}
   \label{fig:NLL30GHz}
    \caption{\Ac{LLF} at 30 GHz carrier for delay only and combined delay and carrier phase observations.}
    \label{fig:NLL30GHz}
\end{figure}

\begin{figure}[t]
    \centering
%
%
\definecolor{mycolor1}{rgb}{0.49400,0.18400,0.55600}%
\begin{tikzpicture}[scale=1\columnwidth/10cm,font=\footnotesize]
\begin{axis}[%
width=8cm,
height=4cm,
scale only axis,
xmode=log,
xmin=0.001,
xmax=1,
xminorticks=true,
xlabel style={font=\color{white!15!black}},
xlabel={bandwidth [GHz]},
ylabel={[m]},
ymode=log,
ymin=0.00001,
ymax=10000,
yminorticks=true,
 legend columns=3,
 legend pos=north east,
axis background/.style={fill=white},
legend style={legend cell align=left, align=left, draw=white!15!black}
]
\addplot [color=blue, line width=1.0pt, dotted]
  table[row sep=crcr]{%
0.0001	157.620713712364\\
0.00016	98.5129460702276\\
0.00026	60.6233514278324\\
0.00042	37.5287413600867\\
0.0007	22.517244816052\\
0.00112	14.0732780100325\\
0.00184	8.56634313654153\\
0.00298	5.28928569504578\\
0.00484	3.25662631637116\\
0.00784	2.0104682871475\\
0.01274	1.23721125362923\\
0.0207	0.761452723248136\\
0.0336	0.469109267001084\\
0.05456	0.288894270000668\\
0.08858	0.177941650160718\\
0.14384	0.109580585172667\\
0.23358	0.0674803980273843\\
0.37926	0.0415600679513696\\
0.61584	0.025594426102943\\
1	0.0157620713712364\\
};
\addlegendentry{$\text{PEB}_{\text{delay}}$}

\addplot [color=black, line width=1.0pt, dashed,mark=o,mark options={solid, black},mark repeat=4]
  table[row sep=crcr]{%
0.0001	0.000162504216949368\\
0.00016	0.000162504216949233\\
0.00026	0.000162504216948871\\
0.00042	0.000162504216947931\\
0.0007	0.000162504216945222\\
0.00112	0.000162504216938621\\
0.00184	0.000162504216920215\\
0.00298	0.000162504216872759\\
0.00484	0.000162504216747139\\
0.00784	0.000162504216418607\\
0.01274	0.000162504215547686\\
0.0207	0.000162504213248804\\
0.0336	0.000162504207199202\\
0.05456	0.000162504191240409\\
0.08858	0.000162504149183991\\
0.14384	0.000162504038261218\\
0.23358	0.000162503745747861\\
0.37926	0.000162502974706814\\
0.61584	0.000162500941582311\\
1	0.000162495581146767\\
};
\addlegendentry{$\text{PEB}_{\text{known}}$}

\addplot [color=red, line width=1.0pt, dashed]
  table[row sep=crcr]{%
0.0001	166.883486089265\\
0.00016	93.1293406760909\\
0.00026	56.9217815305748\\
0.00042	32.3369075404175\\
0.0007	18.7408380910757\\
0.00112	10.5515870554887\\
0.00184	5.29095433623228\\
0.00298	1.97363608501831\\
0.00484	0.913770868538708\\
0.00784	0.539659755905567\\
0.01274	0.240328299191512\\
0.0207	0.0325240660102136\\
0.0336	0.000162504207199203\\
0.05456	0.00016250419124041\\
0.08858	0.000162504149183994\\
0.14384	0.000162504038261216\\
0.23358	0.000162503745747862\\
0.37926	0.000162502974706813\\
0.61584	0.000162500941582312\\
1	0.000162495581146768\\
};
\addlegendentry{$\text{PEB}_{\text{mi}}$}

\addplot [color=blue, line width=1.0pt, mark=square, mark options={solid, blue}]
  table[row sep=crcr]{%
0.0001	245.275452885275\\
0.00016	131.551746679925\\
0.00026	73.7420011359535\\
0.00042	45.2453623543433\\
0.0007	26.9999335000418\\
0.00112	15.1645030766906\\
0.00184	9.24175964489583\\
0.00298	4.96613362666576\\
0.00484	3.41132123127848\\
0.00784	2.17108373802863\\
0.01274	1.16254578911394\\
0.0207	0.754521753712394\\
0.0336	0.464802259910633\\
0.05456	0.30424577965899\\
0.08858	0.199666086338755\\
0.14384	0.114460546375913\\
0.23358	0.0692824706678365\\
0.37926	0.042975781555095\\
0.61584	0.0281272199357815\\
1	0.0162337439352251\\
};
\addlegendentry{$\text{RMSE}_{\text{delay}}$}

\addplot [color=red, line width=1.0pt, mark=o, mark options={solid, red}]
  table[row sep=crcr]{%
0.0001	106.61730467044\\
0.00016	34.296017198645\\
0.00026	80.1697994333883\\
0.00042	67.1141174712427\\
0.0007	23.498784014557\\
0.00112	24.0447285636141\\
0.00184	7.33861781436313\\
0.00298	5.32852954635462\\
0.00484	3.33418259532377\\
0.00784	0.29809206390601\\
0.01274	0.350307182069721\\
0.0207	0.0823\\ 
0.0336	0.000167923959799146\\
0.05456	0.000175593687918269\\
0.08858	0.000158611857121762\\
0.14384	0.000156085889263943\\
0.23358	0.000163133373842509\\
0.37926	0.000160695569823755\\
0.61584	0.000172553237895725\\
1	0.000162459818911118\\
};
\addlegendentry{RMSE \eqref{eq:NLL}}

\addplot [color=black, line width=1.0pt]
  table[row sep=crcr]{%
0.0001	4266.08801971811\\
0.00016	1041.64247655484\\
0.00026	155.846983535885\\
0.00042	79.727204934358\\
0.0007	46.9042978774373\\
0.00112	25.1166187916322\\
0.00184	15.2669881457797\\
0.00298	9.58888186054875\\
0.00484	6.56282702140172\\
0.00784	4.02144943717373\\
0.01274	1.63156487141349\\
0.0207	0.8838\\
0.0336	0.368103070084397\\
0.05456	0.000198815002918428\\
0.08858	0.000163546083647148\\
0.14384	0.000151380920839749\\
0.23358	0.00016427721345202\\
0.37926	0.000159127349809694\\
0.61584	0.000168559916934869\\
1	0.000162694119559173\\
};
\addlegendentry{RMSE \eqref{eq:mixedInteger}}
\end{axis}
\end{tikzpicture}%
    \vspace{-10mm}
    \caption{Impact of bandwidth on PEB  and positioning RMSE.}
    \label{fig:bandwidth}
\end{figure}
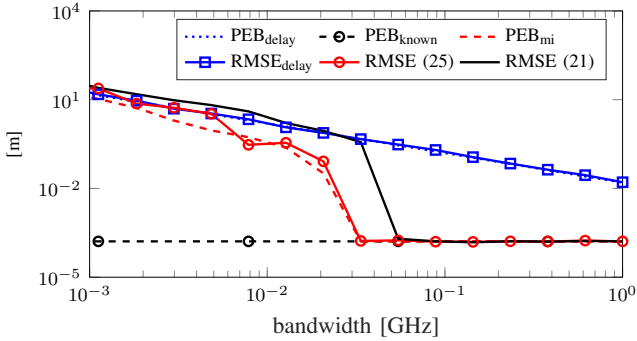

We first vary the carrier frequency $f_c$, leading to the results in Fig.~\ref{fig:carrier}. This figure reveals several interesting facts. The PEB for delay-only estimation ($\text{PEB}_{\text{delay}}$) increases with carrier frequency, due to the increasing path loss. In contrast, the PEB with known ambiguity  ($\text{PEB}_{\text{known}}$) is relatively constant, due to two counter-acting effects: large path loss (leading to lower SNR) vs.~smaller wavelength (leading to better carrier phase estimates for a given SNR). The proposed bound $\text{PEB}_{\text{mi}}$ has a markedly different behavior: for low carrier frequencies (below 20 GHz) it coincides with $\text{PEB}_{\text{known}}$, while for higher carrier frequencies (above 50 GHz), it coincides with $\text{PEB}_{\text{delay}}$. The reason for this will be discussed shortly. Focusing on the algorithms, we observe that both methods show a similar trend as $\text{PEB}_{\text{mi}}$, with the mixed-integer approach performing worse for carriers below 30 GHz, while the directional statistics approach deviates from the bound above 30 GHz.  
To understand the transitions, we consider a 1D cut of the \ac{LLF} at 3 GHz  (Fig.~\ref{fig:NLL3GHz}) and at 30 GHz (Fig.~\ref{fig:NLL30GHz}). At 3 GHz the wavelength (10 cm) is on the order of the uncertainty of the delay-only position estimates. This means that the \ac{LLF} near the delay-only estimate is relatively smooth. Since the  SNR is high enough, there is a clear peak around the true value, separate from the peak of the delay-only \ac{LLF}. 
At 30 GHz, the delay-only \ac{LLF} is much broader (due to the increased path loss), while 
phase estimates in units of cycles also become  worse due to path loss, which means that although the carrier phase accuracy in meter units does not degrade, this accuracy becomes actually worse with respect to the wavelength.
the wavelength has become much smaller. The overall effect on the combined \ac{LLF} is shown. Note that there are many local optima, with \ac{LLF} close to the global optimum. This leads to an increased probability of selecting the wrong optimum, corresponding to an integer ambiguity error.

\subsubsection{Impact of Bandwidth}
In this section, we vary the bandwidth, as shown in Fig.~\ref{fig:bandwidth}. We observe that $\text{PEB}_{\text{delay}}$ improves with bandwidth, as expected, while $\text{PEB}_{\text{known}}$ is approximately constant since the quality of the carrier phase estimates is independent of the bandwidth. The proposed \ac{MICRB} has a different behavior: for small bandwidths, we have poor delay estimation, leading again to a similar effect as shown in Fig.~\ref{fig:NLL30GHz}, so that the bound coincides with the $\text{PEB}_{\text{delay}}$. As the bandwidth increases, the delay estimates get better, while the carrier phase estimates maintain the same quality. Combined, the \ac{MICRB} improves. After a bandwidth of 300 MHz (when the delay estimation is on the order to 10 $\lambda$), the \ac{MICRB}  touches $\text{PEB}_{\text{known}}$, as this case is similar to Fig.~\ref{fig:NLL3GHz}. Fig.~\ref{fig:carrier} and Fig.~\ref{fig:bandwidth} combined clearly show that carrier phase positioning is possible with any bandwidth and any carrier frequency, but not with any arbitrary combination of both. In summary, the bandwidth must be sufficiently large so that the delay estimates are good enough with respect to the wavelength.

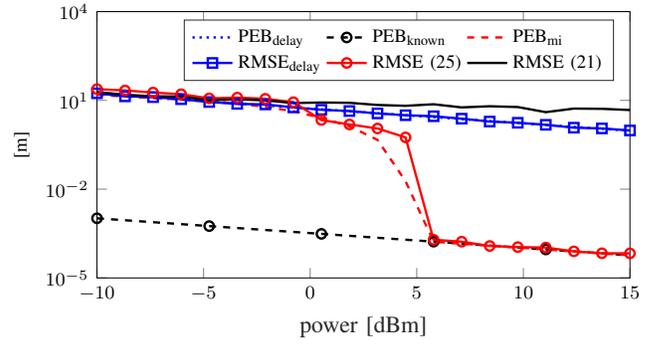
\begin{figure}[t]
    \centering
%
%
\definecolor{mycolor1}{rgb}{0.49400,0.18400,0.55600}%
\definecolor{mycolor2}{rgb}{0.63500,0.07800,0.18400}%
\begin{tikzpicture}[scale=1\columnwidth/10cm,font=\footnotesize]
\begin{axis}[%
width=8cm,
height=4cm,
scale only axis,
xmin=-10,
xmax=15,
xlabel style={font=\color{white!15!black}},
xlabel={power [dBm]},
ylabel={[m]},
ymode=log,
ymin=1e-05,
ymax=1e4,
yminorticks=true,
 legend columns=3,
 legend pos=north east,
axis background/.style={fill=white},
legend style={legend cell align=left, align=left, draw=white!15!black}
]
\addplot [color=blue, line width=1.0pt, dotted]
  table[row sep=crcr]{%
-10	16.7324026468253\\
-8.68421052631579	14.3803293646305\\
-7.36842105263158	12.3588869452941\\
-6.05263157894737	10.6215986194476\\
-4.73684210526316	9.12852085564302\\
-3.42105263157895	7.84532498331624\\
-2.10526315789474	6.74250791197982\\
-0.789473684210526	5.79471379959251\\
0.526315789473685	4.9801510739834\\
1.8421052631579	4.28009140355511\\
3.1578947368421	3.67843909765846\\
4.47368421052632	3.16136103634221\\
5.78947368421053	2.71696862086547\\
7.10526315789474	2.33504443241595\\
8.42105263157895	2.00680731440317\\
9.73684210526316	1.72471047712581\\
11.0526315789474	1.48226798285923\\
12.3684210526316	1.2739056219285\\
13.6842105263158	1.09483275112687\\
15	0.940932147803422\\
};
\addlegendentry{$\text{PEB}_{\text{delay}}$}

\addplot [color=black, line width=1.0pt, dashed,mark=o,mark options={solid, black},mark repeat=4]
  table[row sep=crcr]{%
-10	0.00103504898076972\\
-8.68421052631579	0.000889552180052144\\
-7.36842105263158	0.000764507859760476\\
-6.05263157894737	0.000657041015403146\\
-4.73684210526316	0.00056468078177411\\
-3.42105263157895	0.000485303623106956\\
-2.10526315789474	0.000417084508987156\\
-0.789473684210526	0.000358454953464707\\
0.526315789473685	0.000308066952607299\\
1.8421052631579	0.000264761991350448\\
3.1578947368421	0.000227544407053657\\
4.47368421052632	0.000195558497340606\\
5.78947368421053	0.000168068845889487\\
7.10526315789474	0.000144443413826329\\
8.42105263157895	0.000124139007960601\\
9.73684210526316	0.000106688791750457\\
11.0526315789474	9.16915518511702e-05\\
12.3684210526316	7.88024734645085e-05\\
13.6842105263158	6.77252123969294e-05\\
15	5.82050815482966e-05\\
};
\addlegendentry{$\text{PEB}_{\text{known}}$}

\addplot [color=red, line width=1.0pt, dashed]
  table[row sep=crcr]{%
-10	15.4246015432162\\
-8.68421052631579	13.5154439941246\\
-7.36842105263158	11.9760813216565\\
-6.05263157894737	10.7813305254929\\
-4.73684210526316	9.29571182382971\\
-3.42105263157895	7.86031203304851\\
-2.10526315789474	6.0444457562113\\
-0.789473684210526	4.3073023081607\\
0.526315789473685	2.55572468269006\\
1.8421052631579	1.33417985837054\\
3.1578947368421	0.463388302615758\\
4.47368421052632	0.0185942879115033\\
5.78947368421053	0.000168068845889485\\
7.10526315789474	0.00014444341382633\\
8.42105263157895	0.000124139007960601\\
9.73684210526316	0.000106688791750454\\
11.0526315789474	9.1691551851172e-05\\
12.3684210526316	7.88024734645092e-05\\
13.6842105263158	6.77252123969315e-05\\
15	5.82050815482953e-05\\
};
\addlegendentry{$\text{PEB}_{\text{mi}}$}

\addplot [color=blue, line width=1.0pt, mark=square, mark options={solid, blue}]
  table[row sep=crcr]{%
-10	17.3645073770545\\
-8.68421052631579	13.7004206882168\\
-7.36842105263158	12.8858222411196\\
-6.05263157894737	10.9725563542933\\
-4.73684210526316	8.82602910140839\\
-3.42105263157895	7.62415947382053\\
-2.10526315789474	7.31889750268337\\
-0.789473684210526	5.79076538666724\\
0.526315789473685	4.81925025794006\\
1.8421052631579	4.3655110410638\\
3.1578947368421	3.62894503292705\\
4.47368421052632	3.15450745479285\\
5.78947368421053	2.91133795692812\\
7.10526315789474	2.40947294559919\\
8.42105263157895	1.92742101404\\
9.73684210526316	1.7528812449064\\
11.0526315789474	1.4963601160053\\
12.3684210526316	1.19685016745744\\
13.6842105263158	1.12944962251783\\
15	0.963439423373657\\
};
\addlegendentry{$\text{RMSE}_{\text{delay}}$}

\addplot [color=red, line width=1.0pt, mark=o, mark options={solid, red}]
  table[row sep=crcr]{%
-10	24.0190462881867\\
-8.68421052631579	21.6734731724467\\
-7.36842105263158	18.4127056067363\\
-6.05263157894737	16.1957268425263\\
-4.73684210526316	11.7111604033048\\
-3.42105263157895	12.4570572735914\\
-2.10526315789474	11.3188367302925\\
-0.789473684210526	8.72893679048667\\
0.526315789473685	2.16872750533595\\
1.8421052631579	1.53881885485461\\
3.1578947368421	1.12383743100081\\
4.47368421052632	0.561874313383758\\
5.78947368421053	1.9510e-04\\
7.10526315789474	1.6768e-04\\
8.42105263157895	0.00012\\
9.73684210526316	0.00011\\
11.0526315789474	1.0644e-04\\
12.3684210526316	7.9e-05\\
13.6842105263158	6.8e-05\\
15	6.7567e-05\\
};
\addlegendentry{RMSE \eqref{eq:NLL}}

\addplot [color=black, line width=1.0pt]
  table[row sep=crcr]{%
-10	19.216035958936\\
-8.68421052631579	15.509605122274\\
-7.36842105263158	13.744703857438\\
-6.05263157894737	13.4706193057345\\
-4.73684210526316	10.0959537367024\\
-3.42105263157895	10.8715851002906\\
-2.10526315789474	10.0231832090993\\
-0.789473684210526	8.12209422063347\\
0.526315789473685	8.48843985538991\\
1.8421052631579	8.28310129902789\\
3.1578947368421	7.00543359115543\\
4.47368421052632	6.56101944773631\\
5.78947368421053	7.40300794902578\\
7.10526315789474	5.79586769797794\\
8.42105263157895	6.33355479133611\\
9.73684210526316	5.9670936693671\\
11.0526315789474	4.00114584747551\\
12.3684210526316	5.28445834869614\\
13.6842105263158	5.19268972539256\\
15	4.72603360198602\\
};
\addlegendentry{RMSE \eqref{eq:mixedInteger}}

\end{axis}
\end{tikzpicture}%
    \vspace{-10mm}
    \caption{Impact of transmission power on PEB and positioning RMSE.}
    \label{fig:power}
\end{figure}

\begin{figure}[t]
    \centering
%
%
\definecolor{mycolor1}{rgb}{0.49400,0.18400,0.55600}%
\begin{tikzpicture}[scale=1\columnwidth/10cm,font=\footnotesize]
\begin{axis}[%
width=8cm,
height=4cm,
scale only axis,
xmin=4,
xmax=12,
xminorticks=true,
xlabel style={font=\color{white!15!black}},
xlabel={number of BS},
ylabel={[m]},
ymode=log,
ymin=0.00001,
ymax=1e4,
yminorticks=true,
 legend columns=3,
 legend pos=north east,
axis background/.style={fill=white},
legend style={legend cell align=left, align=left, draw=white!15!black}
]
\addplot [color=blue, line width=1.0pt, dotted]
  table[row sep=crcr]{%
4	15.2176169919147\\
5	10.0913914550671\\
6	5.39339785317973\\
7	5.29125030909978\\
8	4.01452982482993\\
9	4.01447298871339\\
10	3.00559365953909\\
11	2.96598102096157\\
12	2.93889990256433\\
};
\addlegendentry{$\text{PEB}_{\text{delay}}$}

\addplot [color=black, line width=1.0pt, dashed,mark=o,mark options={solid, black},mark repeat=4]
  table[row sep=crcr]{%
4	0.000941345919631802\\
5	0.000624242953064326\\
6	0.000333629967473828\\
7	0.000327311226906815\\
8	0.000248334628992942\\
9	0.000248331113170036\\
10	0.000185922889831015\\
11	0.000183472493312913\\
12	0.000181797283566482\\
};
\addlegendentry{$\text{PEB}_{\text{known}}$}

\addplot [color=red, line width=1.0pt, dashed]
  table[row sep=crcr]{%
4	10.1470352963879\\
5	4.31207004719853\\
6	5.45578242440356\\
7	3.29384454379289\\
8	1.29871527001483\\
9	0.288710673170277\\
10	0.000185922889831008\\
11	0.000183472493312912\\
12	0.000181797283566478\\
};
\addlegendentry{$\text{PEB}_{\text{mi}}$}

\addplot [color=blue, line width=1.0pt, mark=square, mark options={solid, blue}]
  table[row sep=crcr]{%
4	14.7016482069736\\
5	10.3186522144862\\
6	5.36080463974367\\
7	5.40317575305279\\
8	4.07125856657428\\
9	4.2204510300727\\
10	2.8972590001391\\
11	2.74978367692641\\
12	2.84713512865197\\
};
\addlegendentry{$\text{RMSE}_{\text{delay}}$}

\addplot [color=red, line width=1.0pt, mark=o, mark options={solid, red}]
  table[row sep=crcr]{%
4	19.6007869596161\\
5	12.4333161572287\\
6	10.9765013138861\\
7	4.83543537827475\\
8	1.37900218373112\\
9	0.499791604362087\\
10	0.524611712286341\\
11	2.0420e-04\\
12	2.0673e-04\\
};
\addlegendentry{RMSE \eqref{eq:NLL}}

\addplot [color=black, line width=1.0pt]
  table[row sep=crcr]{%
4	15.466615268098\\
5	11.9029814717816\\
6	7.49713818465297\\
7	7.84071276420147\\
8	7.31565352831803\\
9	9.12613330858134\\
10	7.17601189483931\\
11	8.44809059348891\\
12	9.83182718951718\\
};
\addlegendentry{RMSE \eqref{eq:mixedInteger}}
\end{axis}
\end{tikzpicture}%
    \vspace{-10mm}
    \caption{Impact of number of \acp{BS} on PEB  and positioning RMSE.}
    \label{fig:numberofBS}
\end{figure}
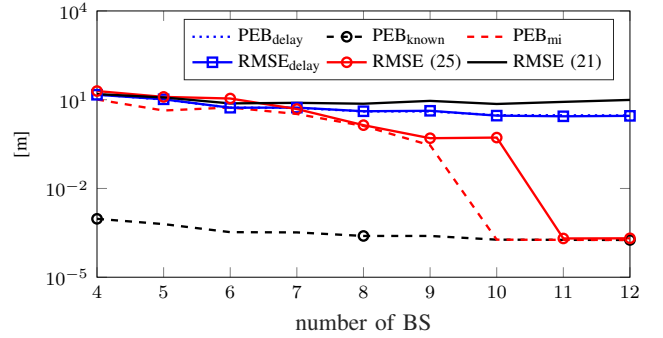

\subsubsection{Impact of Transmission Power}

When we instead change the transmission power (see Fig.~\ref{fig:power}), a different story emerges. Both $\text{PEB}_{\text{delay}}$ and $\text{PEB}_{\text{known}}$ improve with transmission power (due to increased SNR) and lead to roughly parallel curves. The proposed \ac{MICRB} has a thresholding behavior, where for sufficiently large transmit power, a more pronounced global optimum of the carrier phase estimate leads us to correctly identify the correct optimum. We see that the directional statistic approach \eqref{eq:NLL} closely follows the \ac{MICRB}, while the mixed integer approach \eqref{eq:mixedInteger} is unable to improve upon the delay-only estimator. This effect can be ascribed to the poor delay estimates for all considered transmit powers so that the linearization point  $\bm{s}_0$ used in \eqref{eq:linearization} is far away from the \ac{ML} solution. These results clearly indicate that the mixed integer approach requires a good delay-only estimate, while the directional statistics approach does not.

\subsubsection{Impact of Number of BSs}
As a last result, we vary the number of BSs (see Fig.~\ref{fig:numberofBS}). In this figure, BSs are progressively added when computing the bound
$\text{PEB}_{\text{delay}}$ and $\text{PEB}_{\text{known}}$ tend to be relatively flat since performance is dominated by a few BSs with good geometry and SNR. In correspondence to the results as a function of transmission power, we can predict that the mixed-integer algorithm will fail to attain $\text{PEB}_{\text{known}}$, due to poor delay estimates, for the considered number of \acp{BS}. 
The directional statistics approach does not suffer from this shortcoming and closely follows the \ac{MICRB} $\text{PEB}_{\text{mi}}$, harnessing the unique global optimum of the NLL. This shows the great promise of positioning using for instance cell-free deployments, where each \ac{UE} will be surrounded by many phase-coherent access points \cite{fascista2023uplink}.

\section{Conclusions}
This paper studied the \ac{CPP} problem in a cellular context, considering a snapshot positioning scenario under \ac{LoS} conditions. A new fundamental performance bound is derived that can account for the inherent integer ambiguity of the \ac{CPP} positioning problem. The new bound is demonstrated  to be tighter than previously known bounds, at a cost of higher complexity. Results as a function of carrier frequency, bandwidth, transmission power, and the number of \acp{BS} reveal surprising insights as to when carrier phase information can be harnessed. In particular, the standard mixed-integer approach may be far from optimal when the delay estimates are poor. 
{There are several possible extensions of this work, including (i) the presence of  multipath, (ii) a blocked \ac{LoS} path; (iii) reducing the complexity of the \ac{MICRB}.}

{\footnotesize \section*{Acknowledgment}
This work was supported by the Swedish Research Council (VR grants 2018-03701 and 2022-03007) and the  Gigahertz-ChaseOn Bridge Center at Chalmers in a project financed by Chalmers, Ericsson, and Qamcom, by ICREA Academia Program, and by the Spanish R+D project PID2020-118984GB-I00.  The authors are grateful to Erik Agrell for his feedback related to sphere decoding.}
\appendix[FIM per link]\label{sec:AppA}
Consider a generic link and let $\bm{\mu}=\sqrt{E_s}\rho e^{\jmath \vartheta} \bm{d}(\tau)$, then ${\partial \bm{\mu}}/{\partial \tau}  = \sqrt{E_s}\rho e^{\jmath \vartheta} \bm{D} \bm{d}(\tau)$ and ${\partial \bm{\mu}}/{\partial \vartheta}  = \jmath \sqrt{E_s} \rho  e^{\jmath \vartheta} \bm{d}(\tau)$,  
where $\bm{D}=-\jmath 2 \pi  \Delta_f\text{diag}(-(N-1)/2,\ldots,(N-1)/2)$. Since the subcarrier indices are chosen symmetric around 0, then $  J(\tau) ={2}/{N_0}  E_s \rho^2\sum_{n=-(N-1)/2}^{(N-1)/2} (2 \pi n  \Delta_f)^2\approx \text{SNR}\,{2 \pi^2  W^2}/{3}$, using  $\sum_{n=0}^{(N-1)/2}  n^2 \approx N^3 / 24$ for sufficiently large $N$. In addition, $ J(\vartheta)  = {2E_s \rho^2 N}/{N_0}=2\,\text{SNR}$.

\balance 
\bibliographystyle{IEEEtran}
\bibliography{IEEEabrv,reference.bib}
\end{document}